\documentclass[preprint,12pt]{elsarticle}

\usepackage{amssymb}

\def\beq{\begin{eqnarray}}
\def\eeq{\end{eqnarray}}
\def\beqs{\begin{eqnarray*}}
\def\eeqs{\end{eqnarray*}}

\newcommand{{\SD}}{\rm SD}

\newcommand{\lan}{\langle}
\newcommand{\ran}{\rangle}

\newcommand{\be}{\begin{equation}}
\newcommand{\ee}{\end{equation}}
\newcommand{\T}{\mbox{Tr}\> }
\def\centeron#1#2{{\setbox0=\hbox{#1}\setbox1=\hbox{#2}\ifdim
\wd1>\wd0\kern.5\wd1\kern-.5\wd0\fi
\copy0\kern-.5\wd0\kern-.5\wd1\copy1\ifdim\wd0>\wd1
\kern.5\wd0\kern-.5\wd1\fi}}
\def\ltap{\;\centeron{\raise.35ex\hbox{$<$}}{\lower.65ex\hbox{$\sim$}}\;}
\def\gtap{\;\centeron{\raise.35ex\hbox{$>$}}{\lower.65ex\hbox{$\sim$}}\;}
\def\gsim{\mathrel{\gtap}}

\journal{Nuclear Physics B}

\begin{document}

\begin{frontmatter}



\title{Quantum measurements and chiral magnetic effect}


\author{V.I.Shevchenko}

\address{National Research Centre "Kurchatov Institute" \\ ac. Kurchatova sq., 1, Moscow 123182 Russia}

\begin{abstract}
The effect of anisotropy for fluctuations of electric currents in magnetic field
is addressed within framework of quantum measurements theory. It is shown that for free
fermions in uniform magnetic field the anisotropy is of the same sign as one expects for chiral
magnetic effect and is related to triangle anomaly. The corresponding decoherence functional
contains anomalous off-diagonal terms leading to correlation of fluctuations between
observables of opposite P-parity.

\end{abstract}


\end{frontmatter}


\section{Introduction}
\label{}

It is often useful to think about quantum field theory vacuum as about some special medium - {\it aether}.
Despite properties of modern quantum and relativistic aether are quite different from those of its ancient predecessors,
it is important that the set of tools one uses to study them stays more or less intact for centuries: these are either test bodies (from Galilean balls to subatomic particles) or external conditions (heat, pressure etc). In the former case one studies the influence of the medium to test objects movements/interactions, while in the latter one the medium response to external conditions is analyzed. In both approaches properties of the medium are encoded in various correlation functions. There is a huge hierarchy of dynamical scales in Nature characterizing the vacua of different sectors of the Standard Model (and, hopefully, beyond it).

Of particular interest is a question about the fate of discrete symmetries under this or that choice of external conditions. Naively all three main discreet symmetries: charge conjugation ${\bf C}$,  parity reversion ${\bf P}$ and time reversion ${\bf T}$ are not seen in our everyday experience, as manifested by matter over antimatter dominance in the Universe, bio-chirality, arrow of time and numerous other facts. However as is well known the situation is much more subtle. Namely, the interactions governing the macro-world - long ranged gravity and electromagnetism - are invariant under  ${\bf C}$, ${\bf P}$ and ${\bf T}$.
On the other hand, microscopic dynamics respect only the famous ${\bf CPT}$ product: direct  ${\bf P}$-violation is built into electroweak sector of the Standard Model (lefts are doublets and rights are singlets), while conjugate parity ${\bf CP}$ (and hence ${\bf T}$ - invariance) is broken by Cabibbo-Kobayashi-Maskawa mechanism.

Strong interactions stay apart in some sense. Leaving aside strong ${\bf CP}$-problem and all related issues, QCD Lagrangian without $\theta$-term is invariant
under ${\bf C}$-, ${\bf P}$- and ${\bf T}$-transformations. ${\bf C}$--invariance holds at finite temperature but gets broken
at finite density: there is no Furry theorem if some levels in the upper Dirac continuum are occupied. Strong and quite general results \cite{vafa1,vafa2}
guarantee that vacuum expectation value of any local ${\bf P}$-nonconserving observable has to vanish in vector-like theories such as QCD, e.g.
\be
\lan \bar\psi \gamma^5 \psi \ran = 0 \;\;\; ; \;\;\; \lan \T G_{\mu\nu} {\tilde{G}}^{\mu\nu} \ran = 0
\label{vw}
\ee
Despite these results put serious constraints on possible ${\bf P}$-parity violating phenomena in the domain of strong interactions physics, such effects (establishing limits of applicability of (\ref{vw}), in some sense) have been studied for a long time. One can mention T.D.Lee's idea of ${\bf P}$-odd bubbles and A.B.Migdal's hypothesis of pion condensate in nuclei. Closely related effects of $\rho - \pi$ mixing at finite temperature \cite{ioffe} and sphaleron dynamics in QCD \cite{mclerran} were discussed.

Recently the topic has been revitalized in a series of papers \cite{kh1,kh2,kh3,kh4} and numerous subsequent publications. The two general approaches to study any vacuum nicely work together in heavy ion collision experiments with respect to the QCD vacuum. Indeed, test particles used in these experiments --- heavy ions --- are able to create, in the first instants after the collision, highly nontrivial multi-particle state, which itself play a role of external conditions put on the QCD vacuum. These include temperature, density and also, as was noticed in \cite{kh1,kh2,kh3,kh4,st}, extremely strong magnetic field, of the order of $(10^3 - 10^4)\> {\mbox{MeV}}^2 $  in about $0.1 - 0.2 \> {\mbox{Fm}}/c$ after the moment of collision. This is the first (and perhaps the only) case in physics where one can study strong and electromagnetic interactions interplay on the same scale, without treating the latter as a weak perturbation. The main qualitative result can be formulated as follows: if by whatever dynamical mechanism there is an excess of quarks of definite chirality inside a fireball, it transforms into electric current flowing along the magnetic field, whose main effect is charge asymmetry of final particles distribution between upper and lower (with respect to the interaction plane) hemispheres. On quantitative level, for free massless spinors with charge $e$, chemical potentials $\mu_L , \mu_R$ for left-handed and right-handed ones, respectively, in constant and spatially uniform magnetic field ${\bf B}$  electric current is given by the following expression, known as chiral magnetic effect (CME):
\be
{\bf j} = \frac{e^2}{2\pi^2} \mu_5 {\bf B} \;\;\; ; \;\; \mu_5 = \frac{\mu_R - \mu_L}{2}
\label{m}
\ee
The expression (\ref{m}) first explicitly obtained in \cite{vil} (not in heavy ion collision context) is a robust theoretical result and can be reproduced, besides direct computation of the corresponding Feynman diagram, in many complementary ways (Chern-Simons electrodynamics \cite{khchs},
linear response theory \cite{khw},  counting of the number
of zero modes for chiral fermions interacting the external magnetic
field \cite{nielsen}, relativistic hydrodynamics \cite{hydro} etc).
The expression (\ref{m}) is a concrete realization of more general form
\be
j_\mu = {\tilde F}_{\mu\nu} \partial^\nu \phi
\label{m1}
\ee
where dual field strength ${\tilde F}_{\mu\nu} = \frac12 \epsilon_{\mu\nu\alpha\beta} F^{\alpha\beta} $.
One may think of various scenarios corresponding to physical content of the field $\phi$, in particular, in hydrodynamics context \cite{we3}. What is however special and important about the result (\ref{m}) is that proportionality coefficient there is universal and fixed by the famous triangle anomaly \cite{adler,bj}.

The phenomenon of existence of non-dissipative current like (\ref{m}), (\ref{m1}) is also known in electroweak sector \cite{shap}.
Above critical temperature the relevant gauge fields are chiral hypercharge ones, i.e $U(1)_{em} \to U(1)_Y $ at $T>T_c$, those dynamics is different from that of the vector electromagnetic fields in the broken symmetry phase (at $T<T_c$). The hypercharge current ${\bf j_Y}$ can flow along the hypercharge magnetic field ${\bf H_Y}$. In (\ref{m}) both the current and the magnetic field are not chiral but usual vector fields, and the ultimate reason for the effect is generation of nonzero $\mu_5$. As has already been mentioned, in the original picture \cite{kh1,kh2,kh3,kh4} the strong interactions triangle anomaly in the divergence of singlet axial vector current (not to be confused with the abelian anomaly) is supposed to be responsible for it. Thus CME in QCD context should be a subtle interplay of abelian and nonabelian anomalies, like it is the case for $\eta'$ meson, where the latter anomaly is responsible for its mass, while the former one - for its decay to two photons.

There are a few open questions left by the result (\ref{m}). The first concerns physical mechanism of getting effectively nonzero $\mu_5$ - it is worth remembering that there is no such thing as $\mu_L \neq \mu_R$ in the fundamental Lagrangian. The second question is about quantum meaning of (\ref{m}) with respect to quantum field theoretical correlators $\lan \>{\bf j} \>\ran$, $\lan \>{\bf j} \>{\bf j}\>\ran$ etc. Last but not least, it is by far not clear how to extract from the result (\ref{m}) concrete predictions for charge asymmetries and corresponding correlators \cite{voloshin} in real heavy ion collision experiments. An attempt to address some of those questions was undertaken in \cite{we1}.

In the present paper we take a somewhat different prospective and look at the problem using methods of quantum theory of measurements. Throughout the paper $\hbar=c=1$, the Minkowski metric tensor is $g_{\mu\nu}={\mbox{diag}}(1,-1,-1,-1)$ and constant uniform magnetic field is chosen in the third direction: $F_{12} = B$. Also the notation $p^\| = (p^0, 0, 0, p^3)$, $p^\bot = (0, p^1, p^2, 0)$, $p^\| \gamma^\|  = p^0\gamma^0 - p^3 \gamma^3$, $p^\bot \gamma^\bot = p^1\gamma^1 + p^2 \gamma^2$ is used for four-vector components parallel and perpendicular to the field and their products.

\section{CME and triangle anomaly}

It is worth noting that one can easily mimic the effect under discussion without direct use of massless degrees of freedom and any reference to the corresponding anomalies. By way of example let us consider effective Heisenberg-Euler type Lagrangian\footnote{In this model example only two terms are kept in it.} for photon-gluon interaction:
\be
L = - {\frac14 F_{\mu\nu} F^{\mu\nu} } + \frac{\xi(t)}{4} F_{\mu\nu} {\tilde F}^{\mu\nu}  G^a_{\alpha\beta} {\tilde G}^{a\alpha\beta}
\ee
where $F_{\mu\nu} $ stays for the photon and $G^a_{\alpha\beta}$ - for the gluon fields. The factor $\xi(t)$ represents some kind of external potential and encodes external conditions. They are supposed to be time-dependent but slow: ${\dot\xi} / \xi \ll m$.  For example, one could think of time-dependent mass of the heavy fermions integrated out. Corresponding equation of motion reads:
\be
\partial_\mu \left( F^{\mu\nu} +  \xi(t) \cdot G^a_{\alpha\beta} {\tilde G}^{a\alpha\beta} \cdot {\tilde F}^{\mu\nu} \right) = j^\nu
\ee
and for constant fields one gets the following expression for the chiral current
\be
{\bf j}_\chi = \dot{\xi}(t) \cdot G^a_{\alpha\beta} {\tilde G}^{a\alpha\beta} \cdot {\bf B}
\label{he}
\ee
This simple example clearly shows that the crucial feature of non-dissipative currents like (\ref{m}) or (\ref{he}) is not the abelian anomaly but non-stationarity, i.e. time dependence of the corresponding effective Lagrangian. In case of (\ref{m}) it is nothing but the theta-term: $\mu_5 \propto \dot{\theta}(t)$.

From quantum field theoretical point of view a manifestation of the same triangle anomaly in external field (and CME) is given by nonzero correlator of vector and axial vector currents. The latter vanishes in the vacuum:
\be
\Pi^5_{\mu\nu}(k) = i\int d^4 x \> e^{ik(x-y)} \> \lan T\{ j_\mu(x) j^5_\nu(y)\} \ran  = 0
\ee
but does not in external field. For constant abelian field (unit electric charge is assumed in expressions (\ref{p5})-(\ref{sd})) the result is known in the literature (\cite{shub}, see also \cite{gy}) :
\be
\Pi^5_{\mu\nu} (k,F) = \frac{1}{16\pi^2} \left( c_{\|} \cdot [ \tilde F k ]_{\mu\nu}^\|
+  c_\bot \cdot [ \tilde F k ]_{\mu\nu}^\bot \right)
\label{p5}
\ee
where
\be
 [ \tilde F k ]_{\mu\nu}^{\|} = {\tilde F}_{\mu\nu} k^{\|} + {\tilde F}_{\mu\rho} k^\rho k^{\|}_\nu + {\tilde F}_{\nu\rho} k^\rho  k^{\|}_\mu
 \ee
 and analogously for $[ \tilde F k ]_{\mu\nu}^\bot$. Exact form of the functions $c_{\|}$, $c_\bot$ can be found in \cite{shub}. Using these expressions it is easy to check that vector current is conserved as it should be, while the leading term for correlator of axial current divergence is fixed by the triangle anomaly in the massless limit:
\be
 k^\nu \Pi^5_{\mu\nu} (k,F) = \frac{{\tilde F}_{\mu\rho} k^\rho }{2\pi^2}[1 + ...]
\label{sd}
\ee
where the dots in the r.h.s of (\ref{sd}) stay for non-anomalous terms (formally IR-divergent in the massless limit).

According to the famous theorems \cite{bardeen} one would expect non-renormalization of the leading anomaly-driven term in (\ref{sd}) with respect to the strong interactions. In the context of CME this is counter-intuitive in some sense (this was discussed in another context in \cite{we1}). Without external gauge field both  $j^5_\nu$ and $j_\mu$ fluctuates (thermally or quantum-mechanically) in completely uncorrelated way. There are no however fluctuations of  $\partial j^5$ and $\partial j$ in the massless limit since both currents are conserved. With nonzero external field the pattern changes and there are some parts of fluctuations of $j^5_\nu$, $j_\mu$ which become correlated. The amount of this correlation is fixed by the electromagnetic anomaly independently from concrete (strong interaction) physics\footnote{Up to trivial factor $N_c \T Q^2$, taking into account sums over colors and flavors of quarks.} beyond the fluctuation pattern of $j^5_\nu$ and $\partial j^5$.

It is instructive to compare (\ref{sd}) with the chiral magnetic effect interpretation in terms of the standard polarization operator $\Pi_{\mu\nu} (k)$ for vector currents \cite{we1}. It the latter case the term ${\tilde F}_{\mu\rho}k^\rho {\tilde F}_{\nu\sigma}k^\sigma \cdot \tilde\pi^{(F)}(k)$ in $\Pi_{\mu\nu} (k)$ responsible for the discussed effect (i.e. enhanced vector current fluctuations along the magnetic field) is roughly speaking eq.(\ref{sd}) squared, i.e. a product of two triangle anomalies is hidden in imaginary part of the form-factor $\tilde\pi^{(F)}(k)$. On the other hand at large $m$ limit $\tilde\pi^{(F)}(k)$ reproduces the coefficient in front of $(F{\tilde F})^2/m^4$ term of the corresponding Heisenberg-Euler Lagrangian.
It is a remarkable power of topological phenomena such as triangle anomaly to leave traces even in the large mass limit.

\section{Current fluctuation asymmetry and quantum measurements}

The simplest case of free massless fermions in external magnetic field at finite temperature is addressed in this section.
We are interested to study how their fluctuations (both thermal and quantum) are affected by external field. In principle, this information is encoded
in polarization operator $\Pi_{\mu\nu}(x,y) = i\lan T\{ j_{\mu}(x) j_\nu(y) \} \ran$ and we refer an interested reader to \cite{we1} for discussion which structures of  $\Pi_{\mu\nu}(x,y)$ in magnetic field correspond to chiral magnetic effect. In what follows we take a different attitude and consider instead the standard Unruh--DeWitt detector coupled to the current. The corresponding Hamiltonian reads:
\be
H = \int\limits_{\tau_0}^{\tau_1} d\tau \> \mu(\tau) n^\mu  \bar\psi(x(\tau)) \gamma_\mu \psi(x(\tau))
\label{detector}
\ee
Here $x(\tau)$ parameterizes the detector's world-line, $\tau$ - proper time along it, $n^\mu$ - constant vector, fixing a direction the current is measured in, and $\mu(\tau)$ - internal quantum variable of the detector whose evolution in $\tau$ is described by the standard two-level Hamiltonian with the levels $E_0$ and $E_1$, $E_1 - E_0 = \omega > 0$.
An amplitude for the detector to "click" is given by
\be
{\cal A} = i \int\limits_{\tau_0}^{\tau_1} d\tau \> \lan 1 | \mu(\tau) | 0 \ran \cdot \lan \Omega | n^\mu j_\mu(x(\tau)) | \Omega_0 \ran
\ee
where $j_\mu(x(\tau)) = \bar\psi(x(\tau)) \gamma_\mu \psi(x(\tau))$ and $| \Omega_0 \ran$ stays for initial (thermal vacuum) state of the field sub-system, while $| \Omega \ran$ represents final (after the measurement) state.
The corresponding response function reads:
\be
{\cal F}(\omega) \equiv n^\mu n^\nu {\cal F}_{\mu\nu}(\omega) = n^\mu n^\nu \int\limits_{\tau_0}^{\tau_1} d\tau \int\limits_{\tau_0}^{\tau_1} d\tau' \> e^{-i\omega (\tau - \tau')} \cdot   G^+_{\mu\nu}(\tau - \tau')
\ee
where
\be
G^+_{\mu\nu}(\tau - \tau') = \lan \Omega_0 | j_\mu (x(\tau)) j_\nu (x(\tau')) | \Omega_0 \ran
\label{wi}
\ee
Usually one is interested  in detector excitation rate in unit time. For infinite observation time range ($\tau_0 \to -\infty$, $\tau_1 \to \infty$) it is determined by the power spectrum of the corresponding Wightman function:
\be
\dot{{\cal F}}(\omega) = \int\limits_{-\infty}^{\infty}  d s \>  e^{-i\omega s} \>  G^+(s)
\label{eri}
\ee
The details of this standard procedure can be found in \cite{brag,bd}, see also \cite{bessa} in the context of vector current measurements.

To compute (\ref{eri}) it is convenient to use the exact fermion propagator in external magnetic field given by \cite{schw1}
\be
S(x,y) = e^{i\phi(x,y)} \int \frac{d^4 p}{(2\pi)^4} e^{ip(x-y)} {\tilde S}(p)
\ee
where the gauge-dependent phase $\phi(x,y)$ is irrelevant for gauge-invariant quantities and Fourier transfrom ${\tilde S}(p)$ reads:
\be
{\tilde S}(p) = \int\limits_0^\infty \> du \> e^{iu(p_0^2 - p_3^2 - p_\bot^2 \frac{\tan(qBu)}{qBu} - m^2)} \left[ P_0(p^{\|}, p^\bot) + P_1 (p^{\|}) \right]
\label{hf}
\ee
where
\be
P_0(p^\| , p^\bot) = p^\| \gamma^\| - p^\bot \gamma^\bot (1+\tan^2(qBu)) +m
\ee
\be
P_1(p^\|) = (p^\|\gamma^\| + m)\gamma^1 \gamma^2 \tan(qBu)
 \ee
  and $q$ stays for quark electric charge. We take in the rest of the paper $m=0$. Notice that the result for anomalous correlator (\ref{sd}) comes from the interference of the above tensor structures $P_0(p^\| , p^\bot)$ and $P_1(p^\|)$. Diagrammatically it corresponds to the standard triangle diagram for abelian anomaly. We will see that the tensor structure $P_1(p^\|)$ is also responsible for charge fluctuations asymmetry.

It is customary in quantum measurements theory to compare response functions of a given detector in a state of inertial movement versus some non-inertial one. We are interested in another kind of asymmetry, namely between the detector oriented to measure current along the magnetic field direction and perpendicular to it. This choice is fixed by the vector $n_\mu = (0,{\bf n})$. With respect to its spatial movement the detector is supposed to be always at rest, so we can take $x(\tau) = (\tau, 0,0,0)$. Therefore it is convenient to switch to the coordinate space (as in (\ref{eri}) we denote $s=\tau-\tau'$):
\be
S(s) = \frac{is\gamma^0}{32\pi^2} \int\limits_0^\infty \> \frac{du}{u^3}  \left( \frac{qBu}{\tan(qBu)} + \gamma^1 \gamma^2 qBu \right)  e^{- i\frac{s^2}{4u}}
\label{opk}
\ee
As is stressed above the leading term in (\ref{sd}) gets its first power of the field just from the second term in the brackets in (\ref{opk}), which in its turn has come from the term $P_1 (p^{\|})$ in (\ref{hf}). The same term also produces angular asymmetry in (\ref{eri}). The response function asymmetry given by $\delta {\dot{\cal F}}(\omega) =  {\dot{\cal F}}_{33}(\omega) - ({\dot{\cal F}}_{11}(\omega) + {\dot{\cal F}}_{22} (\omega))/2$ is quadratic in $B$ for all values of the magnetic field.\footnote{Notice that ${\dot{\cal F}}_{11}(\omega) = {\dot{\cal F}}_{22}(\omega)$ for our choice of the field along the third axis.}
Explicitly, one gets:
\be
G^+_{33}(s) = - \left[ \frac{s}{16\pi^2} \int\limits_0^\infty \> \frac{du}{u^3}  \> \frac{qBu}{\tan(qBu)} e^{- i\frac{s^2}{4u}} \right]^2  - \frac{(qB)^2}{16\pi^4 s^2}
\label{g33}
\ee
\be
G^+_{11}(s)  = - \left[ \frac{s}{16\pi^2} \int\limits_0^\infty \> \frac{du}{u^3}  \> \frac{qBu}{\tan(qBu)} e^{- i\frac{s^2}{4u}} \right]^2  + \frac{(qB)^2}{16\pi^4 s^2}
\label{g11}
\ee
and $G^+_{11}(s) = G^+_{22}(s)$. These results are exact for free fermions in external magnetic field in the massless limit.

To compute the response function one needs to take into account $s \to s - i\epsilon$ prescription corresponding to definition of the Wightman function (\ref{wi}) and switch on the temperature introducing sum over periodic shifts in imaginary time with $\beta = 1/kT$ and Fermi-Dirac statistics factor $(-1)^k$ for fermions (see, e.g. \cite{brag,bd}):
\be
\delta  {\dot{\cal F}}(\omega)  = - \frac{(qB)^2}{8\pi^4} \int\limits_{-\infty}^{+\infty} ds \> e^{-i\omega s} \left[\sum\limits_{k=-\infty}^{+\infty} \frac{(-1)^k}{(s - i\epsilon + ik\beta)} \right]^2
\label{eq2w}
\ee
where we denoted $\delta  {\dot{\cal F}}(\omega)  = {\dot{\cal F}}_{33}(\omega) - {\dot{\cal F}}_{11}(\omega)$

Taking into account that $
\sum\limits_{k=-\infty}^{\infty} \frac{(-1)^k}{x+ik} = \frac{\pi}{\sinh \pi x}
$
and doing the integral with the help of residues (see, e.g. \cite{ps}):
\be
\int\limits_{-\infty}^\infty  \frac{e^{-i\omega s} ds}{\sinh^{2n}(s-i\epsilon)} = \frac{(-1)^n}{(2n-1)!} \left( \frac{2\pi}{\omega}\right) \frac{1}{e^{\pi\omega} - 1} \prod\limits_{l=1}^n \left(\omega^2 + 4(n-l)^2\right)
\ee
one gets
\be
\delta  {\dot{\cal F}}(\omega)  =  \frac{(qB)^2}{4\pi^3}  \frac{\omega}{e^{\beta \omega} - 1}
\label{mm}
\ee
Expression (\ref{mm}) is the main result of this section. It is positive, which corresponds to the fact that the detector measuring the current along magnetic field clicks more often than measuring perpendicular currents. It is also worth noticing the change of statistics from Fermi-Dirac to Bose-Einstein - what is relevant is the statistic of operators whose fluctuations are being measured by the detector (Bose-currents in our case) and not the statistics of primary fluctuating fields.

The fact that current fluctuations are suppressed in perpendicular direction is obvious from general physics: the charged particle moving in the orthogonal plane is deflected by the magnetic field (or, using quantum mechanical language, confined to Landau levels). What is less obvious is that fluctuations along the field are enhanced (exactly by the same amount), since classically (i.e. neglecting spin effects) magnetic field has no influence on a charge moving in parallel direction. This enhancement is caused by spin interaction with the magnetic field and, to our view, can legally be called a particular case of CME.

It is instructive to compare (\ref{mm}) with fluctuation pattern without magnetic field. The latter can easily be obtained from (\ref{g33}) or (\ref{g11}) putting $B=0$. The result reads:
\be
{\dot{\cal F}}^{(0)}(\omega) =  \frac{1}{60\pi^3}  \frac{\omega}{e^{\beta \omega} - 1} \left( \omega^2 + 4\left(\frac{2\pi}{\beta}\right)^2 \right) \left( \omega^2 + \left(\frac{2\pi}{\beta}\right)^2 \right)
\label{mmfree}
\ee
 The ratio is maximal for $\omega \to 0$ but even in this region it is rather small numerically:
 \be
 \frac{\delta {\dot{\cal F}}(0)}{{\dot{\cal F}}^{(0)}(0)} =\frac{15}{64\pi^4} \cdot \frac{(qB)^2}{T^4} \approx 0.0024 \cdot \frac{(qB)^2}{T^4}
 \label{gf}
 \ee

At large magnetic fields the fluctuations are frozen in $1-2$ plane, so that
\be
G^+_{11}(s)  \to 0 \;\; ; \;\;
G^+_{33}(s)  =  - \frac{(qB)^2}{8\pi^4 s^2}
\label{ikj1}
\ee
and correspondingly
\be
\frac{{\dot{\cal F}}_{33}(\omega) - {\dot{\cal F}}_{11}(\omega)}{{\dot{\cal F}}_{33}(\omega) + {\dot{\cal F}}_{11}(\omega)} \to 1
\label{ikj2}
\ee

The pattern (\ref{ikj1}), (\ref{ikj2}) is easy to understand in term of the corresponding energy-momentum tensor. Indeed, at zero magnetic field one has the standard thermal pressure for massless fermions
\be
T_{11} = T_{22} = T_{33} = \frac{7 \pi^2 T^4}{180}
\ee
which is isotropic. At large magnetic field, however, all the pressure is along the magnetic field and there is no pressure in orthogonal directions:
\be
T_{11}=T_{22} \to 0 \;\; ; \;\; T_{33} = \frac{qB T^2}{12}
\ee
It is interesting to note that for strong but slowly varying magnetic field the plasma as a whole is to experience buoyancy force in the direction of the field gradient:
\be
f_3 = - \int\limits_V d^3 x \frac{\partial T_{33}}{\partial z}
\ee
This effect can be called magnetic Archimedes force. Since in real scattering events the fields are indeed highly inhomogeneous, this effect could be important  for such phenomena as charge dependence of the elliptic flow etc. These questions are to be explored in future research.

 Despite detailed quantitative analysis of the correspondence between the above results and experimentally
  measured quantities is beyond the scope of the present paper, we find it instructive to make a few comments on this point. The most important outcome of each heavy ion collision event is final particles distribution with respect to azimuthal angle:
 \be \frac{dN_{\pm}}{d\phi} = \frac{N_\pm}{2\pi} \left[ 1+2v_{1,\pm}\cos\Delta\phi + 2v_{2,\pm} \cos2\Delta\phi + 2a_{\pm} \sin\Delta\phi
+ ... \right] \label{ne} \ee
where a reaction plane by definition is taken to lie at zero angle: $\Delta \phi = \phi - \Psi_{RP}$.
 One can also compute averages of various functions over many events with the distribution (\ref{ne}) as a weight.  In particular, the commonly accepted probe for the asymmetry with respect to the reaction plane is given in terms of event average of the cosine \cite{voloshin},
\be \lan
\cos(\Delta \phi_\alpha + \Delta {\phi_\beta}' ) \rangle_e
\ee
where the final particles electric charge is indicated by $\alpha, \beta = +, -$. The most recent experimental results for this quantity can be found in \cite{alice}. They clearly indicate presence of nonzero
{\bf P}-even averages $\lan a_\alpha a_\beta \rangle_e$ and hence nonzero {\bf P}-odd correlations proportional to $a_+ , a_-$ coefficients at least in some events.

It is clear that the dynamics described by (\ref{ne}) is different from the physics discussed above in the following important respect. The fluctuations in (electrically neutral as a whole) thermal plasma are isotropic, while the distribution (\ref{ne}) is obviously not, even without any external field. The dominant
isotropic term in (\ref{ne}) corresponds to zero total current since equal amounts of same charged particles flow in all directions. The asymmetries in the currents along the magnetic field are encoded in the terms proportional to $a_\pm $, while those in the currents in orthogonal direction - by the directed flow coefficients $v_{1,\pm}$.
To be more precise, one is to study averages $\lan (a_+ - a_-)^2 \rangle_e$ and $\lan (v_{1,+} - v_{1,-})^2 \rangle_e$ corresponding to electric current fluctuations along the magnetic field and in perpendicular azimuthal direction, respectively.

This means that if one takes a (very crude) assumption that orthogonal to the reaction plane magnetic field is the only source of the asymmetry,  one can make qualitative conclusion from the picture considered above that
\be
\lan (a_+ - a_-)^2 \rangle_e \gsim \lan (v_{1,+} - v_{1,-})^2 \rangle_e
\label{az}
\ee
and
\be
\frac{\lan (a_+ - a_-)^2 \rangle_e - \lan (v_{1,+} - v_{1,-})^2 \rangle_e}{
\lan (a_+ - a_-)^2 \rangle_e + \lan (v_{1,+} - v_{1,-})^2 \rangle_e} \propto B^2
\label{az2}
\ee

It is worth reminding that in conventional CME picture there is {enhancement} of the current along the magnetic field, but {\it no suppression} of the current in orthogonal directions, contrary to our picture where both effects are present and they are precisely equal in magnitude. At the moment the current status of experimental data \cite{alice}
does not allow to draw definite conclusion about validity of the pattern (\ref{az}), (\ref{az2}) and further studies are needed. Coming back to (\ref{gf}) one can reasonably conclude that for not too large $qB/T^2$ ratio relative asymmetry for free massless fermion thermal fluctuations can be at most of per cent level.

As the last remark of this section let us notice that one can view the discussed asymmetry in another way: via nonlocal "order parameters".
Consider nonlocal current of the following form:
\be
{\cal J}^5_\mu(x-y) = \lan \bar\psi(y) \exp\left(iq\int\limits_x^y A_{\alpha}(z) dz^\alpha \right) \gamma_\mu \gamma^5 \psi(x) \ran
\ee
Physically it corresponds to Green's function of a system composed of massless spinor ("light quark") and infinitely heavy "antiquark" (static source).
By construction it is gauge-invariant and identically vanishes at zero external field. If the system is supposed to be static, it means that it is created at the point $y=(\tau',0,0,0)$ with the quantum numbers of a scalar, propagates in external magnetic field and annihilates at the point $x=(\tau, 0,0,0)$ with the quantum numbers of axial vector. The nontrivial effect of anomaly is that this Green's function is nonzero for the third  (i.e. parallel to the field) component of the axial vector. Indeed, one easily gets ($s=\tau-\tau'$):
\be
{\cal J}^5_0(s) = {\cal J}^5_1(s) = {\cal J}^5_2(s) = 0 \;\; ; \;\; {\cal J}^5_3(s) = \frac{1}{2\pi^2} \frac{qB}{s}
\label{uh}
\ee
As is (\ref{opk}), the above result is exact in the field $B$. There are well developed phenomenological methods to study this kind of objects in QCD where quarks interact non-perturbatively with quantum gluon field \cite{stj}. It would be interesting to check (non)renormalization of the result (\ref{uh}) in genuine nonperturbative domain.

\section{Parity violation and decoherence}

Discussing possible parity-violating effects in quantum mechanical context it is useful to remember that
if parity is conserved in the sense that $[{\bf P},H]=0$, any ${\bf P}$-odd quantity has to vanish after a non-selective measurement.
Indeed, after quantization the eigenstates $\left|\psi_n\right\ran$ can be chosen as states of definite - even and odd - parity according to
\be
\left|\psi_+\right\ran  = {\bf P}  \left|\psi_+\right\ran \;\; ; \; \left|\psi_-\right\ran  = - {\bf P}  \left|\psi_-\right\ran
\ee
Now consider expectation value of some real {\bf P}-odd operator $O_-$ in a state
$ \left|\psi\right\ran$ having no definite parity, for example,
\be
\left|\psi\right\ran = c_+ \left|\psi_+\right\ran + c_- \left|\psi_-\right\ran
\ee
It is proportional to transition matrix element between the states of opposite parity:
\be
\left\lan O_- \right\ran = \left\lan\psi\right| O_- \left|\psi\right\ran = 2 \mbox{Re} \left\{c_+^* c_- \left\lan\psi_+\right| O_- \left|\psi_-\right\ran \right\}
\ee
Let us now couple the system in question with the environment causing its decoherence.
 After decoherence the density matrix $\rho =  \left|\psi\right\ran \left\lan \psi \right| $ loses its off-diagonal terms and becomes:
 \be
 \rho \to \rho_d =  |c_+|^2 \left|\psi_+ \right\ran \left\lan \psi_+ \right| +  |c_-|^2 \left|\psi_- \right\ran \left\lan \psi_- \right|
 \ee
 and average of {\bf P}-odd operator $O_-$ vanishes:
\be
\left\lan O_- \right\ran_d = {\mbox{Tr}} \> \rho_d \cdot O_- = 0
\ee

On the other hand, if one is monitoring some ${\bf P}$-odd observable, it can lead to nonzero result for measurement of correlated ${\bf P}$-odd quantity. The simplest way to see it is to use a language of decoherence functionals (\cite{feynmanvernon}, see also \cite{hu}) and path integral formalism.
Generally, for some filter function $\alpha[\Phi] $ the amplitude is given by
\be
\Psi[\alpha] = \int {\cal D} {\Phi} \> \alpha[\Phi] \> e^{iS[\Phi]}
\ee
To illustrate the point on quantum-mechanical example, consider three-dimensional system given by Lagrangian $ L = {{\dot q}^2} - V(q)$,
where $q=(x,y,z)$ and the potential $V(q)$ is invariant under ${\bf P}$-parity transformation: $V(q) = V(-q)$, but not invariant under separate reflections $x \to -x$ or $y \to -y$ or $z \to -z$. Suppose that the system is open to external observer who is monitoring the $y$-coordinate continuously in time. As is well known to describe this situation in path integral representation one has to introduce quantum corridor and perform the shift:
\be
\int {\cal D} y(t) \to  \int {\cal D} y(t) \> \exp\left(-\kappa \int\limits_0^T (y(t) - {\bar y}(t) )^2 dt \right) \label{shift}
\ee
where the corridor width is given by $\Delta y \propto (\kappa T)^{-1/2}$.
Then the transition amplitude takes the standard form up to the measure shift (\ref{shift}):
$$
U(q''; q') =
$$
\be = \int {\cal D}p \int {\cal D} q \> \exp\left(\frac{i}{\hbar} \int\limits_0^T (p{\dot q} - H(p,q)) dt -\kappa \int\limits_0^T (y(t) - {\bar y}(t) )^2 dt \right)
\label{pol}
\ee
where we denote $q=(x,y,z), p=(p_x , p_y, p_z)$.
Consequently, all amplitudes and correlators computed with (\ref{pol}) become dependent on the function $ {\bar y}(t) $ (which has the meaning of continuous observation result) and quantum corridor width $\Delta y $. For example, for $x$-coordinate one would have
\be
\lan x (T) \ran = X[{\bar y}(t), \kappa]
\label{x2}
\ee
where the functional $X[{\bar y}(t), \kappa]$ depends, generally speaking, on the function ${\bar y}(t)$ in all past moments of time and vanishes at the point $\kappa = 0$, corresponding to no measurement: $X[{\bar y}(t), 0] = 0$. Its exact form of course depends on the potential $V(q)$ and is of no importance for us at the moment. What is crucial is the fact that monitoring ${\bf P}$-odd quantity (coordinate $y$ in our example) can result in nonzero quantum average for some other ${\bf P}$-odd quantity (coordinate $x$), despite the interaction is still strictly ${\bf P}$-even. One can say that ${\bf P}$-parity is broken "event-by-event" by measuring apparatus.

Coming back to our discussion of CME in the previous sections it is clear that crucial missing ingredient is of course the fact that in strong interaction domain the singlet axial vector current is not conserved because of triangle nonabelian anomaly:
\be
\partial^\nu j^5_\nu(x) = -\eta(x) = -\frac{g^2 N_f}{16\pi^2}G_{\alpha\beta}^a (x) {\tilde{G}}^{a\alpha\beta}(x)
\label{eta2}
\ee
We are interested to find common distribution for the vector current and some ${\bf P}$-odd quantity, which we have chosen in this section to be the field $\eta(x)$ from (\ref{eta2}). The corresponding amplitude reads:
\be
\Psi[\lambda , \kappa] = \int {\cal D} {\bar\psi} {\cal D} \psi {\cal D} A_\mu \> e^{iS_{QCD} + i\int dx \lambda(x)  n_\mu j^\mu(x) + i\int dx \kappa(x) \eta(x)}
\label{olk}
\ee
 The vector current is given by the standard expression $j_\mu = \bar\psi Q \gamma_\mu \psi$, where $Q$ is quark electric charges diagonal matrix in flavor space. The closed-time-path functional is given by
\be
e^{iW[\lambda,\kappa ; \lambda' , \kappa']} = \Psi[\lambda , \kappa] \Psi^*[\lambda' , \kappa']
\ee
and the mean current is
\be
\lan n_\mu j^\mu (x) \ran[\lambda, \kappa] = -i \left.\frac{\delta}{\delta \lambda(x)}e^{iW[\lambda,\kappa ; \lambda' , \kappa']} \right|_{\kappa = \kappa' \atop \lambda = \lambda'}
\ee
It is a functional of  ${\bf P}$-even field $\lambda(x)$ and ${\bf P}$-odd field $\kappa(x)$ in the same sense as $\lan x (T) \ran$ from (\ref{x2}) is a functional of $\bar{y}(t)$.

It is easy to compute $\Psi[\lambda , \kappa] $ in Gaussian approximation. It reads:
\be
\Psi_{Gauss}[\lambda , \kappa] = e^{\frac{i}{2} \int dp (\lambda(p), \kappa(p)) {D(p)} (\lambda(-p), \kappa(-p))^{{T}} }
\ee
where
\be
D(p) = \left( \begin{array}{cc}\Pi(p) & \Delta(p) \\ \Delta(p) & \Pi^5(p) \\ \end{array} \right)
\ee
with the components
\begin{equation}
\Pi(p)  = i \int dx \> e^{ipx} \lan T\{ j_\mu(x) j_\nu(0) \} \ran \> n^\mu n^\nu
\ee
\be
\Pi^5(p) = i \int dx \> e^{ipx} \> \lan T\{ \eta(x) \eta(0) \} \ran \\
\ee
\be
\Delta(p) = \frac{e^2}{2\pi^2} n^\mu p^\alpha {\tilde{F}}_{\alpha\mu} \cdot N_c \mbox{Tr} Q^2
\end{equation}
The non-diagonal terms of the matrix $D(p)$ arise due to correlation of fluctuations of the quantities of opposite ${\bf P}$-parity in external abelian field.

To make the above picture suitable for concrete computations let us take a model profile for the $\kappa$-field, corresponding to the measurement taking place inside 3-dimensional "decoherence volume" $V \sim R^3$ for the time period $\tau$ starting from the moment $t=0$. We chose Gaussian Ansatz
\be
\kappa(t, {\bf x}) = \kappa \cdot f(t, {\bf x}) = \kappa \cdot
\exp\left( -{\bf x}^2 / 2R^2 \right)\> \exp\left( -{t}^2 / 2\tau^2 \right)
\label{theta1}
\ee
It leads to the following expression for the current parallel to magnetic field:
\be
\lan j_3(t, {\bf x}) \ran [0,\kappa]= -\kappa B \> \cdot \left(\frac{N_c \mbox{Tr} Q^2}{2\pi^2} \right)\cdot \left( \frac{t  \cdot f(t, {\bf x})}{\tau^2} \right)  \cdot e^{-\int dp \> \kappa(p) \Im \{ \Pi^5(p) \} \kappa(-p) }
\label{fg}
\ee
where we switched off the ${\bf P}$-even filter ($\lambda = 0$), but has kept the  ${\bf P}$-odd one. The above expression is a generalization of (\ref{m}). The current is linear both in $\kappa$ and in $B$ and vanishes being integrated over $\kappa$ in symmetric limits. Notice that due to the form-factor the current flows only inside the volume (where the measurement has been done).

Expression (\ref{fg}) remarkably demonstrates two different faces of decoherence (see, e.g \cite{breuer}). On one hand, the mere existence of this quantum current is due to classical nature of the field $\kappa$. On the other hand, the last exponent in the right hand side of (\ref{fg}) is responsible for current damping due to decoherence. This is analogous to measuring electric current by applying external electromagnetic field. If the field is weak and/or momentum of its dominant Fourier modes $k$ is small, linear response theory works well and one gets conductivity as proportionality coefficient between applied field and induced current via Green-Kubo formula. If however the quanta of the field have $k^2 > 4m^2$, they decay into real charged particles with the mass $m$ instead of being absorbed by charged current carriers and the linear response picture is not valid anymore. Quantitatively this process is controlled by the discussed factor with the imaginary part of the corresponding polarization operator in the exponent.

The relevance of this exponential damping depends on particular form of $\Im \{ \Pi^5(p) \} $. It is clear, for example, that the presence of mass gap, i.e. if $\Im \{ \Pi^5(p) \} = 0 $ for $p^2 < m_*^2$ makes this effect exponentially small for $\tau m_*  \gg 1$ since $\kappa(p)$ falls off exponentially for large $p$. In massless case, on the other hand, the suppression can be important since it causes a maximum value for $\kappa$ - roughly speaking, impossibility to apply "too strong" classical field to the system, in some analogy with a famous Schwinger's picture of electron-positron pairs creation in strong electric field.

On the other hand, maximizing (\ref{fg}) with respect to $t, {\bf x}$ and $\kappa$, and using the notation from (\ref{m}) we come to expression for maximal possible effective $\mu_5$ in our model (taking into account three quark flavors):
\be
\mu_5^{eff} = \frac{0.17}{\tau\sqrt{\zeta(\tau, R)}}
\ee
where dimensionless factor $\zeta(\tau, R)$ is given by
\be
\zeta(\tau, R) = (2\pi)^4 (R^3 \tau)^2 \int \frac{d^4 p}{(2\pi)^4} \> \Im \{ \Pi^5(p) \}\> \exp\left( -{\bf p}^2 R^2  - {p_0}^2 \tau^2 \right)
\ee

Actual value of $\zeta(\tau, R)$ in finite temperature case is determined by interplay of three length scales: $R, \> c\tau , \> {(\alpha_s T)}^{-1}$. For them being of the same order one would expect $\zeta(\tau, R) \sim {\cal O}(1)$ on dimensional grounds. Thus $\mu_5^{eff}$ is given parametrically by the inverse measurement time scale $\tau$. This can be rephrased using the language of energy-time uncertainty principle: $\mu_5^{eff}$ has a meaning of energy difference between left and right handed fermions; one expects to have this quantity of the order $\tau^{-1}$ if the measurement took time $\tau$. It is physically clear, however, that at times larger than the life time of strong magnetic field, which does not exceed 0.2-0.3 Fm$/c$ \cite{st}, the above picture has no sense because there is no magnetic field. Thus we estimate maximal effective $\mu_5^{eff}$ in the ballpark of 100-200 MeV for realistic heavy ions collision conditions.

There is no commonly accepted way to relate the above results for the current and final particle distributions (\ref{ne}) in quantitative way. The problem is that the event average $  \lan (a_+ - a_-)^2 \rangle_e$
related to the current fluctuations orthogonal to the reaction plane is, obviously, always positive, even without any magnetic field. True asymmetry of interest is encoded in the ratios like (\ref{az2}), which, however, require independent knowledge of quantities $\lan (v_{1,+} - v_{1,-})^2 \rangle_e$, having nothing to do with CME.
Thus a systematic way to proceed could be to compute thermal averages of higher order terms $\lan j_1^2 \rangle$, $\lan j_3^2 \rangle$ with realistic time-dependent magnetic field and pseudoscalar filter field. There will be magnetic field-dependent contribution to the latter quantity given by (\ref{fg}) squared (compare with \cite{fkh}). Then one would expect
\be
\frac{\lan (a_+ - a_-)^2 \rangle_e - \lan (v_{1,+} - v_{1,-})^2 \rangle_e}{
\lan (a_+ - a_-)^2 \rangle_e + \lan (v_{1,+} - v_{1,-})^2 \rangle_e} = \frac{\lan j_3^2 \rangle - \lan j_1^2 \rangle}{\lan j_3^2 \rangle + \lan j_1^2 \rangle}
\label{az3}
\ee
This study is a subject of future work.

It is instructive to discuss the physical meaning of the $\kappa$ field from another prospective. Like in (\ref{shift}), this field encodes classical dynamics of a measuring apparatus responsible for a measurement process. In quantum mechanical example above it is the width of a corresponding quantum corridor. Of course experimentally this corridor must be realized as some boundaries, external potentials or alike. One can imagine "event by event" experimental setup where this parameter fluctuates according to its own distribution. What is important however is that in a given experimental "event" it assumes to take some particular value (controlled or measured by some other device), and perhaps some different value in another event later in time and so on. This is principally different from quantum case where a picture of some quantum observable "fluctuating in time" is just wrong - all quantum histories coexist simultaneously at any given moment until the measurement is done.

Coming back to quantum field theoretical context, one could think of the classical pseudoscalar field $\kappa(y)$ in (\ref{olk}) as just the standard $\theta$-term introducing {\bf CP}-violation by hand. In fact however its meaning is different. This field corresponds to nothing but {\it classicalization} of some $0^-$-degrees of freedom in the problem. By no means one should think about them only as instantons or any other classical configurations. Correspondingly, (\ref{fg}) is to be understood in distributional sense: if in a given event some pseudoscalar degrees of freedom have been classicalized (and quantitatively the amount of this is encoded in $\kappa(y)$) then there is the electric current in the system given by (\ref{fg}).

The above formalism is of course unable to provide a detailed microscopic picture of such classicalization. However one can rely on general arguments from Color Glass Condensate theory (see e.g. \cite{mcl} and references therein). The key point is a transition from quantum fluctuations to classical (and hence space-time dependent) ones at large occupation numbers. In terms of the field $\eta(x)$ (\ref{eta2}) this corresponds to transition from genuine quantum (vacuum) case when $\lan \eta(x) \rangle = 0$ identically for any $x$ (local {\bf P}-parity conservation) to classically fluctuating (with some distribution) field $\eta_{cl}(x) \neq 0$ with globally conserved {\bf P}-parity $\int dx \> \eta_{cl}(x) = 0$.

\section{Conclusion}

We discussed the phenomenon of CME using the ideas of quantum measurements theory. First, it was shown that in the simplest case of free massless fermions the nonzero asymmetry of electric current fluctuations in magnetic field is detected by the standard Unruh--DeWitt detector at rest. This asymmetry (\ref{mm}) is of desired sign (i.e. detector clicks more often measuring current components along the field than in perpendicular plane) but numerically it is rather weak for usual thermal fluctuations even in its maximum. We find it remarkable that this asymmetry is exactly quadratic in magnetic field $B$ and gets no higher order corrections. Despite the effect is not linear in $B$ as conventional (abelian) triangle anomaly is, its origin can be traced to the same asymmetry in the fermion Green's function in the magnetic field which is responsible for CME in the standard approach. Second, taking into account singlet axial vector current non-conservation due to nonabelian anomaly, we computed the electric current (\ref{fg}) along magnetic field under assumption that particular {\bf P}-odd quantity (dual field to the topological charge density in our example) can be treated as external (classical). This can be understood as a model for chiral chemical potential $\mu_5$ generation in quark-gluon dense and hot medium via decoherence along well known line of thought about classical features of intense gluon field in this system.

{\bf Acknowledgements}

The author acknowledges useful discussions with A.Andrianov, A.Gorsky, D.Kharzeev, M.Polikarpov and V.Zakharov on various topics of strong interaction physics in external abelian fields.

\bibliographystyle{elsarticle-num}

\end{document}